\documentstyle[aps,preprint,tighten,epsf]{revtex}

\begin{document}

\draft

\title{On the effectiveness of Gamow's method for 
calculating decay rates}
\author{R. M. Cavalcanti\footnote{Present address: 
Instituto de F\'\i sica, Universidade de S\~ao Paulo, 
Cx. Postal 66318, CEP 05315-970, S\~ao Paulo, SP.
E-mail: rmoritz@fma.if.usp.br}}
\address{Departamento de F\'\i sica, Pontif\'\i cia Universidade
Cat\'olica do Rio de Janeiro, \\
Cx. Postal 38071, CEP 22452-970, Rio de Janeiro, RJ} 
\author{C. A. A. de Carvalho\footnote{E-mail: aragao@if.ufrj.br}}
\address{Instituto de F\'\i sica,
Universidade Federal do Rio de Janeiro, \\
Cx. Postal 68528, CEP 21945-970, Rio de Janeiro, RJ}
%\date{\today}
\maketitle

\begin{abstract}

We examine Gamow's method for calculating the decay rate
of a wave function initially located within a potential well.
Using elementary techniques, we examine a very simple, 
exactly solvable model, in order to show why it is so reliable 
for calculating decay rates, in spite of its conceptual problems. 
We also discuss the regime of validity of the exponential decay law.

\end{abstract}


%\begin{multicols}{2}

%%%%%%%%%%%%%%%%%%%%%%%%%%%%%%%%%%%%%%%%%%%%%%%%%%%%%%%%%%%%%%%%%%%%%%%%%

\section{Introduction}
\label{I}

Complex-energy ``eigenfunctions'' made their d\'ebut in Quantum
Mechanics through the hands of Gamow, in the theory of alpha-decay
\cite{Gamow}. Gamow imposed an ``outgoing wave boundary condition''
on the solutions of the Schr\"odinger
equation for an alpha-particle trapped in the nucleus. 
Since there is only an {\em outgoing} flux of alpha-particles, 
the wavefunction $\psi(r,t)$ must behave far from the nucleus as 
(for simplicity, we consider an s-wave, and use units such that 
$\hbar=2m=1$)
\begin{equation}
\psi(r,t)\sim\frac{e^{-iEt+ikr}}{r}\qquad(r\rightarrow\infty).
\end{equation}
This boundary condition, together with the requirement
of finiteness of the wave function at the origin, gives rise
to a quantization condition on the values of $k$ (and,
therefore, on the values of $E=k^2$). It turns out
that such values are complex:
\begin{equation}
k_n=\kappa_n-iK_n/2,\qquad E_n=\epsilon_n-i\Gamma_n/2,
\end{equation}
and so it follows that
\begin{equation}
\label{|psi|}
|\psi_n(r,t)|^2\sim\frac{e^{-\Gamma_nt+K_nr}}{r^2}
\qquad(r\rightarrow\infty).
\end{equation}
Thus, if $\Gamma_n>0$, the probability of finding the alpha-particle
in the nucleus decays exponentially in time. 
The lifetime of the nucleus is then given by $\tau_n=1/\Gamma_n$, 
and the energy
of the emitted alpha-particle by $\epsilon_n$.

Although very natural, this interpretation suffers from
some difficulties. How can the energy, which is an observable
quantity,
be complex? (In other words, how can the Hamiltonian, which is a
Hermitean operator, have complex eigenvalues?) 
Also, these ``eigenfunctions'' are not normalizable, 
since $\Gamma_n$ positive implies $K_n$ positive and
so, according to (\ref{|psi|}),
$|\psi_n(r,t)|^2$ diverges exponentially with $r$.

In spite of such problems (which, in fact, are closely related),
it is a fact of life that alpha-decay, as well as other
types of decay, does obey an exponential decay law and, in many cases,
Gamow's method provides a very good estimate for the decay rate. 
Why this method works is a question that has been addressed
in the literature using a variety of techniques 
\cite{Beck,Nuss,CSM,FGR,Skib,BGB,Moshinsky,Holstein}.
Here we examine this question in a very elementary way,
using techniques that can be found in any standard
quantum mechanics textbook and some rudiments of complex analysis. 

This paper is structured as follows. In Section \ref{II},
we show Gamow's method in action for a very simple potential. Some
of the results obtained there are used in Section \ref{III}, where we
study the time evolution of a wave packet initially confined
in the potential well defined in Section \ref{II}. This is done
with the help of the propagator, built with the {\em true}
eigenfunctions (i.e., associated to {\em real} eigenenergies) 
of the Hamiltonian. As a bonus, we show that the
exponential decay law is not valid either for very small or for 
very large times. This is the content of Section \ref{IV}, where
the region of validity of the exponential decay law is roughly
delimited.

%%%%%%%%%%%%%%%%%%%%%%%%%%%%%%%%%%%%%%%%%%%%%%%%%%%%%%%%%%%%%%%%%%%%%%%

\section{Decaying States}
\label{II}

In order to exhibit Gamow's method in action, we shall study
the escape of a particle from the potential well given by:
\begin{equation}
\label{V}
V(x)=\cases{(\lambda/a)\,\delta(x-a) & for $x>0$, \cr
+\infty & for $x<0$. \cr}
\end{equation}
(Escape from this potential well was studied in detail
in Refs.\cite{Nuss,Moshinsky,Winter}. In this section we
follow closely the treatment of Ref.\cite{Nuss}.)
The positive dimensionless constant $\lambda$
is a measure of the ``opacity'' of the barrier at $x=a$; in the limit
$\lambda\rightarrow\infty$, the barrier becomes impenetrable,
and the energy levels inside the well are quantized. If $\lambda$
is finite, but large, a particle is not confined to the
well anymore, but it usually stays there for a long time before it
escapes.
If $\lambda$ is small, the particle can easily
tunnel through the barrier, and quickly escape from
the potential well. Metastability, therefore, can only be achieved
if the barrier is very opaque, i.e., $\lambda\gg 1$.
Since this is the most interesting situation,
we shall assume this to be the case in what follows.

To find out how fast the particle escapes from the potential
well, we must solve the Schr\"odinger equation 
\begin{equation}
i\,\frac{\partial}{\partial t}\,\psi(x,t)
=-\frac{\partial^2}{\partial x^2}\,\psi(x,t)
+\frac{\lambda}{a}\,\delta(x-a)\,\psi(x,t).
\end{equation}
$\psi(x,t)=\exp(-iEt)\,\varphi(x)$ is a particular solution
of this equation, provided $\varphi(x)$ satisfies the time-independent
Schr\"odinger equation
\begin{equation}
\label{Sch}
-\frac{d^2}{d x^2}\,\varphi(x)+\frac{\lambda}{a}\,\delta(x-a)\,
\varphi(x)=E\,\varphi(x).
\end{equation}
Denoting the regions $0<x<a$ and $x>a$ by the indices 1 and 2,
respectively, the corresponding wave functions $\varphi_j(x)$
$(j=1,2)$
satisfy the free-particle Schr\"odinger equation:
\begin{equation}
\label{free}
-\frac{d^2}{d x^2}\,\varphi_j(x)=E\,\varphi_j(x).
\end{equation}
Since the wall at the origin is impenetrable, $\varphi_1(0)$ must
be zero; the solution of Eq.\ (\ref{free}) which obeys this boundary
condition is
\begin{equation}
\varphi_1(x)=A\,\sin kx\qquad(k=\sqrt{E}\,).
\end{equation}
To determine $\varphi_2(x)$, we follow Gamow's reasoning 
\cite{Gamow,BGB,Gold}
and require $\varphi_2(x)$ to be an outgoing wave.
Therefore, we select, from the admissible solutions of 
Eq.\ (\ref{free}),
\begin{equation}
\label{phi2}
\varphi_2(x)=B\,e^{ikx}.
\end{equation}
The wave function must be continuous at $x=a$, so that 
$\varphi_1(a)=\varphi_2(a)$, or
\begin{equation}
\label{cont1}
\frac{B}{A}=e^{-ika}\,\sin ka.
\end{equation}
On the other hand, the derivative of the wave function has a
discontinuity at $x=a$, which can be determined by integrating both
sides of (\ref{Sch}) from $a-\varepsilon$ to $a+\varepsilon$, with
$\varepsilon\rightarrow 0^+$:
\begin{equation}
\label{discont1}
\varphi_2'(a)-\varphi_1'(a)=\frac{\lambda}{a}\,\varphi_2(a),
\end{equation}
from which there follows another relation between $A$ and $B$:
\begin{equation}
\label{discont2}
\frac{B}{A}=-\frac{ka\,e^{-ika}\,\cos ka}{\lambda-ika}.
\end{equation}
Combining (\ref{cont1}) and (\ref{discont2}), we obtain a quantization
condition for $k$:
\begin{equation}
\label{quant}
ka\,{\rm cotan}\,ka=-\lambda+ika.
\end{equation}
The roots of Eq.\ (\ref{quant}) are complex and situated
in the half-plane ${\rm Im}\,k<0$; when $\lambda\gg 1$, those 
which are closest to the origin
are given by \cite{Nuss,Moshinsky}
\begin{equation}
\label{poles}
k_na \approx \frac{n\pi\lambda}{1+\lambda}
-i\left(\frac{n\pi}{\lambda}\right)^2
\quad(n=1,2,\ldots;n\pi\ll\lambda).
\end{equation}
(For each one of these roots, which are located in the fourth 
quadrant of the complex $k$-plane, there is a corresponding one 
in the third quadrant, given by $-k_n^*$. The latter are associated 
to ``growing states'' \cite{Nuss} and play no role in what follows.)
The corresponding eigenenergies are
\begin{equation}
E_n = k_n^2\approx\left(\frac{n\pi}{a}\right)^2
-i\,\frac{2(n\pi)^3}{(\lambda a)^2}. 
\end{equation}
The imaginary part of $E_n$ gives rise to an exponential decay
of $|\psi_n(x,t)|^2$, with lifetime equal to
\begin{equation}
\label{tau}
\tau_n=1/\Gamma_n\approx\frac{(\lambda a)^2}{4(n\pi)^3}.
\end{equation}
Since the corresponding value of $B/A$ is very small 
($\sim n/\lambda$),
one may be tempted to say that the probability of finding the particle
outside the well is negligible in comparison with the probability
of finding the particle inside the well. Normalizing $\psi_n$ in 
such a way that the latter equals one when $t=0$, the probability
of finding the particle inside the well at time $t$, if it were in the
$n$-th decaying state at $t=0$, would be
\begin{equation}
P_n(t)=\int_0^a|\psi_n(x,t)|^2\,dx=\exp(-\Gamma_nt).
\end{equation}

The trouble with this interpretation is that 
${\rm Im}\,k_n\equiv-K_n/2<0$,
and so $\psi_n(x,t)$ diverges exponentially as 
$x\rightarrow\infty$, since,
according to (\ref{phi2}), 
\begin{equation}
|\psi_n(x,t)|^2=|B_n|^2\,\exp(-\Gamma_nt+K_nx)
\end{equation}
outside the well. Because of this ``exponential catastrophe'', the
decaying states are nonnormalizible and, therefore, cannot be
accepted as legitimate solutions of the Schr\"odinger equation 
(although one can find in the literature \cite{Legget}
the assertion that they are ``rigorous'' solutions of the 
time-dependent Schr\"odinger equation). 

%%%%%%%%%%%%%%%%%%%%%%%%%%%%%%%%%%%%%%%%%%%%%%%%%%%%%%%%%%%%%%%%%%%%%%%%

\section{Time Evolution of a Wave Packet}
\label{III}

We now return to Eq.\ (\ref{free}) and write, 
for the solution in region 2,
instead of (\ref{phi2}), the sum of an outgoing plus an incoming wave:
\begin{equation}
\varphi_2(x)=e^{-ikx}+B\,e^{ikx}.
\end{equation}
Continuity of the wave function at $x=a$ implies
\begin{equation}
\label{cont2}
A\,\sin ka=e^{-ika}+B\,e^{ika}.
\end{equation}
As before, the derivative of the wave function has a discontinuity
at $x=a$, given by Eq.\ (\ref{discont1}), from which it follows, 
instead of (\ref{discont2}),
\begin{equation}
\label{discont3}
kA\,\cos ka=-\left(\frac{\lambda}{a}+ik\right)e^{-ika}
-\left(\frac{\lambda}{a}-ik\right)B\,e^{ika}.
\end{equation}
Solving (\ref{cont2}) and (\ref{discont3}) for $A$ and $B$, we find
\begin{mathletters}
\label{AB}
\begin{eqnarray}
A(k)&=&-\frac{2ika}{ka+\lambda\,e^{ika}\,\sin ka}, 
\\
B(k)&=&-\frac{ka+\lambda\,e^{-ika}\,\sin ka}
{ka+\lambda\,e^{ika}\,\sin ka}.
\end{eqnarray}
These expressions show a couple of interesting features: 
\begin{itemize}

\item $|B|=1$ for
real values of $k$, implying a zero net flux of probability through
$x=a$; therefore, unlike the solution found
in the previous section, there is no loss or accumulation of 
probability in the well region. 

\item $|A|\ll 1$ if $ka\ll\lambda$, except if $k$ is close to
a pole of $A(k)$, in which case $|A|$ may become very large. 

\end{itemize}
To find the poles of $A$ we must solve the equation 
$A(k)^{-1}=0$, which, after some algebraic manipulations, reads
\end{mathletters}
\begin{equation}
ka\,{\rm cotan}\,ka=-\lambda+ika.
\end{equation}
This is the same as Eq.\ (\ref{quant})! Is this a coincidence? 
In fact, no.
According to (\ref{AB}), $A$ and $B$ have the same poles; 
therefore, near a pole both $|A|$ and $|B|$ are
very large, and Eqs.\ (\ref{cont2}) and (\ref{discont3}) become
equivalent to Eqs.\ (\ref{cont1}) and (\ref{discont2}), respectively.
In what follows, we shall show that the poles of $A$ (and $B$)
play an important role in the decay process.

Suppose that at $t=0$ the particle is known to be in
the region $x<a$ with probability 1; in other words, its wave function
$\psi(x,0)$ is zero outside the well. Then, at a later time $t$,
the wave function is given by
\begin{equation}
\label{psi1}
\psi(x,t)=\int_0^{a}G(x,x';t)\,\psi(x',0)\,dx',
\end{equation}
where the propagator, $G(x,x';t)$, can be written as
\begin{equation}
\label{G}G(x,x';t)=\int_0^{\infty}e^{-ik^2t}\,
\varphi_k^{}(x)\,\varphi_k^{*}(x')\,dk.
\end{equation}
The function $\varphi_k(x)$ is the solution of Eq.\ (\ref{Sch})
corresponding to the energy $E=k^2$:
\begin{equation}
\label{varphi}
\varphi_k(x)=\frac{1}{\sqrt{2\pi}}\times
\cases{A(k)\,\sin kx & for $x<a$, \cr
e^{-ikx}+B(k)\,e^{ikx} & for $x>a$. \cr}
\end{equation}
With this normalization, these functions satisfy the completeness
relation \cite{Landau}
\begin{equation}
\label{completeness}
\int_0^{\infty}\varphi_k^{}(x)\,\varphi_k^{*}(x')\,dk=\delta(x-x').
\end{equation}
Eqs.~(\ref{psi1})--(\ref{varphi}) give, for $x<a$,
\begin{equation}
\label{psi2}\psi(x,t)=\frac{1}{2\pi}\int_0^{\infty}e^{-ik^2t}\,
\phi(k)\,|A(k)|^2\,\sin kx\,dk,
\end{equation}
where
\begin{equation}
\phi(k)\equiv\int_0^a \psi(x',0)\,\sin kx'\,dx'.
\end{equation}
It is clear that the integral over $k$ is dominated by the
resonances, i.e., the neighborhood of the poles of $A(k)$.

Since, for $t>0$, $e^{-ik^2t}\to 0$ when $|k|\to\infty$ 
in the fourth quadrant, one can 
rotate\footnote{For this to be possible
$\phi(k)$ must be an analytic function of $k$, but this can
be shown to be the case \cite{Watson} if $\psi(x',0)$ is 
continuous in $[0,a]$.}
%Here we are assuming that 
%$\phi(k)$ is an analitic function of $k$, but this can be proved
%This is a reasonable assumption, as 
%$$
%\phi(k)=(-1)^n\,\sqrt{2a}\,\frac{n\pi\,\sin ka}{k^2a^2-n^2\pi^2}
%$$
%for $\psi(x,0)=\sqrt{2/a}\,\sin (n\pi x/a)$ $(n=1,2,\ldots)$,
%which form a basis for functions with support in $[0,a]$.}
the integration contour by $45^{\rm o}$ in the clockwise sense 
(see Fig.~1), thus obtaining
\begin{equation}
\label{psi3}
\psi(x,t)=e^{-i\pi/4}\int_0^{\infty}e^{-k^2t}
\,f(e^{-i\pi/4}\,k,x)\,dk+\sum_{n=1}^{\infty}C(k_n,x)\,e^{-ik_n^2t},
\end{equation}
where
\begin{equation}
f(k,x)\equiv\frac{1}{2\pi}\,\phi(k)\,|A(k)|^2\,\sin kx
\end{equation}
and
\begin{equation}
C(k_n,x)=-2\pi i\,\lim_{k\to k_n}\,(k-k_n)\,f(k,x).
\end{equation}
The sum in (\ref{psi3}) takes into account the poles of $A(k)$ which
are
situated in the region $-\pi/4<{\rm arg}\,k<0$, and it corresponds
to an expansion in Gamow states (for $x<a$). 

Let us put aside, for a moment, the integral in (\ref{psi3})
(it will be discussed in the next section).
Then, the ``nonescape'' probability (i.e., the probability of 
finding the particle inside the well) is given by
\begin{equation}
\label{P(t)}
P(t)=\int_0^a|\psi(x,t)|^2\,dx
\approx\sum_{n=1}^{\infty}c_n\,e^{-\Gamma_nt}
+\,{\rm interference\, terms},
\end{equation}
where $c_n\equiv\int_0^a|C(k_n,x)|^2\,dx$. For $\lambda\gg 1$,
the interference terms are usually negligible, for $k_n\approx n\pi/a$
and, therefore, the functions $C(k_n,x)\propto\sin k_nx$ are 
approximately orthogonal.
On the other hand, since the decay rate $\Gamma_n$ of 
the $n$-th decaying
mode is a rapidly increasing function of $n$
($\Gamma_n\approx n^3\,\Gamma_1$), the decay becomes almost a pure
exponential one when $\Gamma_1t\gtrsim 1$. The system,
therefore, ``loses memory'' of the initial state. 

Finally, let us note that no exponential catastrophe occurs
with $\psi(x,t)$. In fact, one can easily show, using
(\ref{psi1}), (\ref{G}), (\ref{completeness}) and the
orthogonality of the eigenfunctions $\varphi_k(x)$, that
\begin{equation}
\int_0^{\infty}|\psi(x,t)|^2\,dx=\int_0^{\infty}|\psi(x,0)|^2\,dx,
\end{equation}
so that an exponential growth of $|\psi(x,t)|^2$ is completely
ruled out.

%%%%%%%%%%%%%%%%%%%%%%%%%%%%%%%%%%%%%%%%%%%%%%%%%%%%%%%%%%%%%%%%%%%%%%%%%

\section{Breakdown of Exponential Decay}
\label{IV}

In order to derive expression (\ref{P(t)}) for the
nonescape probability, we had to neglect the first term on
the right hand side of (\ref{psi3}). In this section we
show that such approximation is not valid either for
very small or for very large times. That it cannot be valid for 
very small $t$ follows from the fact   
that initially the decay is
slower than exponential \cite{CSM,FGR}.
This can be easily proved with the
help of the continuity equation \cite{Landau2}:
\begin{equation}
\frac{d}{dt}\,P(t)=-\frac{\hbar}{m}\,{\rm Im}\left[\psi(x,t)\,
\frac{\partial}{\partial x}\,\psi^*(x,t)\right]_{x=a}.
\end{equation}
Since, by hypothesis, $\psi(a,0)=0$, it follows that $dP/dt=0$ when
$t=0$,
whereas for the expression (\ref{P(t)}) one has
$dP/dt\approx-\sum\,c_n\Gamma_n<0$ at $t=0$.  

On the other hand, the exponential decay does not last forever.
After a sufficiently long time, it obeys a power 
law \cite{Beck,Nuss,CSM,FGR,Moshinsky,Patra}.
To see this, note that the integral in
(\ref{psi3}), which we shall denote here by $I(x,t)$, is dominated 
by small values of $k$
when $t\rightarrow\infty$, and so can it be approximated by
\begin{equation}
I(x,t)\approx\frac{e^{-i\pi/4}}{2\pi}\,\phi'(0)\,|A(0)|^2\,x\,
\int_0^{\infty}k^2\,e^{-k^2t}\,dk
\sim\frac{a^{3/2}\,x}{\lambda^2\,t^{3/2}} 
\end{equation}
Therefore, the nonescape probability behaves asymptotically 
as\footnote{Garc\'\i a-Calder\'on, Mateos and Moshinsky
\cite{Moshinsky} argue that the nonescape
probability $P(t)$ decays as $t^{-1}$
when $t\rightarrow\infty$, in contrast to Eq.\ (\ref{Pas}).
However, there is a flaw in their argument; when
properly corrected \cite{Ricardo}, it also leads to
$P(t)\sim t^{-3}$ asymptotically.}
\begin{equation}
\label{Pas}
P(t)\approx\int_0^a|I(x,t)|^2\,dx\sim\frac{a^6}{\lambda^4 t^3}.
\end{equation}
Comparing (\ref{Pas}) with (\ref{P(t)}), and using (\ref{tau}), one
finds
that they become comparable in magnitude when
\begin{equation}
e^{-t/\tau_1}\sim\frac{a^6}{\lambda^4 t^3}
\sim\lambda^{-10}\,\left(\frac{\tau_1}{t}\right)^{3},
\end{equation}
or, since $\lambda\gg 1$, when
\begin{equation}
\frac{t}{\tau_1}\sim 10\,\ln \lambda.
\end{equation}
In practice, when the decay begins to obey a power law
the nonescape probability is so small $(\sim\lambda^{-10})$ 
that it should be very difficult to observe deviations from 
exponential decay. (On the other hand, experimental evidence
for non-exponential decay at {\em small} times has been reported
recently \cite{Stevenson}.)

In closing this section, we would like to remark that
the breakdown of the exponential decay law for either small or 
large times is not a peculiar feature of the potential (\ref{V}).
It is possible to show that an exponential decay cannot
last forever if the Hamiltonian is bounded below
\cite{Khalfin,CSM,FGR}, 
and cannot occur for very small times if, besides that,
the energy expectation value of the initial state is 
finite \cite{CSM,FGR} --- conditions which certainly must be 
satisfied by any realistic Hamiltonian
or state. 

%%%%%%%%%%%%%%%%%%%%%%%%%%%%%%%%%%%%%%%%%%%%%%%%%%%%%%%%%%%%%%%%%%%%%%

\section{Conclusion}

In this paper we showed that decaying states,
although plagued by the exponential catastrophe, give a fairly
good description of the decay of a metastable state, provided
some conditions are satisfied. In fact, the main objective
of this paper was to show that one {\em can} compute the decay rate
solving
the time independent Schr\"odinger equation subject to the ``outgoing
wave boundary condition.'' This is far from
being a trivial result, since the corresponding eigenstates
are unphysical. The effectiveness of the decaying states
in describing the decay may be understood by noticing \cite{Skib} 
that they are good approximate solutions to the 
time-dependent Schr\"odinger
equation, although nonuniform ones (i.e., they are not valid
in the entire range of values of $t$ and $x$). 

%%%%%%%%%%%%%%%%%%%%%%%%%%%%%%%%%%%%%%%%%%%%%%%%%%%%%%%%%%%%%%%%%%%%%%%

\acknowledgments

We thank Gernot Muenster and Pavel Exner, for bringing some references
to our attention, and Gilberto Hollauer, for useful discussions.
This work had financial support from CNPq, FINEP, CAPES and FUJB/UFRJ.

%%%%%%%%%%%%%%%%%%%%%%%%%%%%%%%%%%%%%%%%%%%%%%%%%%%%%%%%%%%%%%%%%%%%%%%

%\end{multicols}

%%%%%%%%%%%%%%%%%%%%%%%%%%%%%%%%%%%%%%%%%%%%%%%%%%%%%%%%%%%%%%%%%%%%%%%%%

\begin{figure}[hbt]
\begin{center}
\leavevmode
\epsfxsize=12cm
\epsfbox{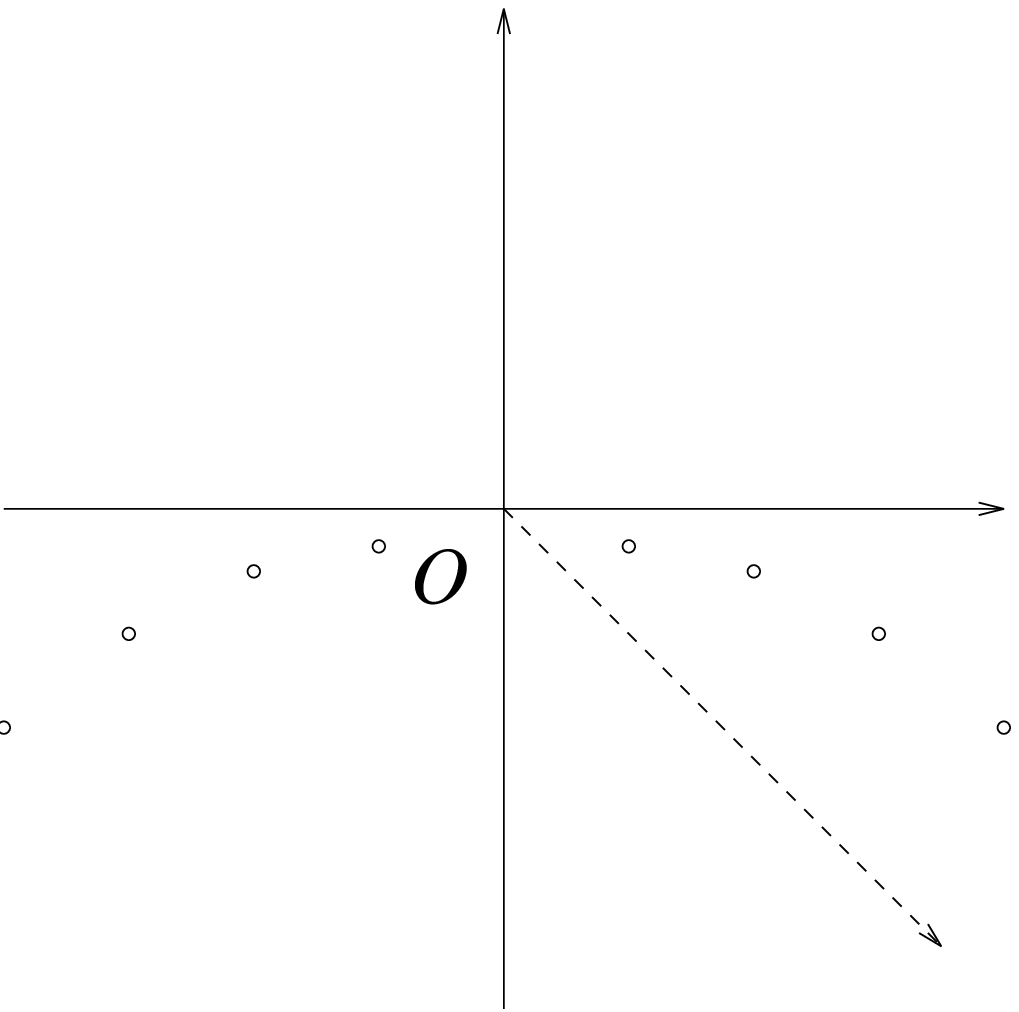}
\end{center}
\end{figure}
FIG.~1. Complex $k$-plane. The poles of $A(k)$ are represented
by the small circles. Those in the fourth quadrant give rise to
the sum over decaying modes in Eq.~(\ref{psi3}) when one rotates
the integration contour of Eq.~(\ref{psi2}) --- the positive real
semi-axis --- by $45^{\rm o}$ in the clockwise sense (dashed line).

%\end{multicols}

\end{document}